\documentstyle[12pt]{article}

\def\datetitle{September 14, 1994}
\def\datehead{\ }

\makeatletter

\@addtoreset{equation}{section}

\begin{document}

    \def\@evenhead{\rm\small\thepage\hfil\datehead}
    \def\@oddhead{\datehead\hfil\rm\small\thepage}

%
%

\makeatother

%
%

\def\bbbr{{\rm I\!R}} 
\def\bbbn{{\rm I\!N}} 
\def\bbbm{{\rm I\!M}}
\def\bbbh{{\rm I\!H}}
\def\bbbf{{\rm I\!F}}
\def\bbbk{{\rm I\!K}}
\def\bbbp{{\rm I\!P}}
\def\bbbone{{\mathchoice {\rm 1\mskip-4mu l} {\rm 1\mskip-4mu l}
{\rm 1\mskip-4.5mu l} {\rm 1\mskip-5mu l}}}
\def\bbbc{{\mathchoice {\setbox0=\hbox{$\displaystyle\rm C$}\hbox{\hbox
to0pt{\kern0.4\wd0\vrule height0.9\ht0\hss}\box0}}
{\setbox0=\hbox{$\textstyle\rm C$}\hbox{\hbox
to0pt{\kern0.4\wd0\vrule height0.9\ht0\hss}\box0}}
{\setbox0=\hbox{$\scriptstyle\rm C$}\hbox{\hbox
to0pt{\kern0.4\wd0\vrule height0.9\ht0\hss}\box0}}
{\setbox0=\hbox{$\scriptscriptstyle\rm C$}\hbox{\hbox
to0pt{\kern0.4\wd0\vrule height0.9\ht0\hss}\box0}}}}
\def\bbbe{{\mathchoice {\setbox0=\hbox{\smalletextfont e}\hbox{\raise
0.1\ht0\hbox to0pt{\kern0.4\wd0\vrule width0.3pt
height0.7\ht0\hss}\box0}}
{\setbox0=\hbox{\smalletextfont e}\hbox{\raise
0.1\ht0\hbox to0pt{\kern0.4\wd0\vrule width0.3pt
height0.7\ht0\hss}\box0}}
{\setbox0=\hbox{\smallescriptfont e}\hbox{\raise
0.1\ht0\hbox to0pt{\kern0.5\wd0\vrule width0.2pt
height0.7\ht0\hss}\box0}}
{\setbox0=\hbox{\smallescriptscriptfont e}\hbox{\raise
0.1\ht0\hbox to0pt{\kern0.4\wd0\vrule width0.2pt
height0.7\ht0\hss}\box0}}}}
\def\bbbq{{\mathchoice {\setbox0=\hbox{$\displaystyle\rm Q$}\hbox{\raise
0.15\ht0\hbox to0pt{\kern0.4\wd0\vrule height0.8\ht0\hss}\box0}}
{\setbox0=\hbox{$\textstyle\rm Q$}\hbox{\raise
0.15\ht0\hbox to0pt{\kern0.4\wd0\vrule height0.8\ht0\hss}\box0}}
{\setbox0=\hbox{$\scriptstyle\rm Q$}\hbox{\raise
0.15\ht0\hbox to0pt{\kern0.4\wd0\vrule height0.7\ht0\hss}\box0}}
{\setbox0=\hbox{$\scriptscriptstyle\rm Q$}\hbox{\raise
0.15\ht0\hbox to0pt{\kern0.4\wd0\vrule height0.7\ht0\hss}\box0}}}}
\def\bbbt{{\mathchoice {\setbox0=\hbox{$\displaystyle\rm
T$}\hbox{\hbox to0pt{\kern0.3\wd0\vrule height0.9\ht0\hss}\box0}}
{\setbox0=\hbox{$\textstyle\rm T$}\hbox{\hbox
to0pt{\kern0.3\wd0\vrule height0.9\ht0\hss}\box0}}
{\setbox0=\hbox{$\scriptstyle\rm T$}\hbox{\hbox
to0pt{\kern0.3\wd0\vrule height0.9\ht0\hss}\box0}}
{\setbox0=\hbox{$\scriptscriptstyle\rm T$}\hbox{\hbox
to0pt{\kern0.3\wd0\vrule height0.9\ht0\hss}\box0}}}}
\def\bbbs{{\mathchoice
{\setbox0=\hbox{$\displaystyle     \rm S$}\hbox{\raise0.5\ht0\hbox
to0pt{\kern0.35\wd0\vrule height0.45\ht0\hss}\hbox
to0pt{\kern0.55\wd0\vrule height0.5\ht0\hss}\box0}}
{\setbox0=\hbox{$\textstyle        \rm S$}\hbox{\raise0.5\ht0\hbox
to0pt{\kern0.35\wd0\vrule height0.45\ht0\hss}\hbox
to0pt{\kern0.55\wd0\vrule height0.5\ht0\hss}\box0}}
{\setbox0=\hbox{$\scriptstyle      \rm S$}\hbox{\raise0.5\ht0\hbox
to0pt{\kern0.35\wd0\vrule height0.45\ht0\hss}\raise0.05\ht0\hbox
to0pt{\kern0.5\wd0\vrule height0.45\ht0\hss}\box0}}
{\setbox0=\hbox{$\scriptscriptstyle\rm S$}\hbox{\raise0.5\ht0\hbox
to0pt{\kern0.4\wd0\vrule height0.45\ht0\hss}\raise0.05\ht0\hbox
to0pt{\kern0.55\wd0\vrule height0.45\ht0\hss}\box0}}}}

%
%

\def\bbbz{{\mathchoice {\hbox{$\sf\textstyle Z\kern-0.4em Z$}}
{\hbox{$\sf\textstyle Z\kern-0.4em Z$}}
{\hbox{$\sf\scriptstyle Z\kern-0.3em Z$}}
{\hbox{$\sf\scriptscriptstyle Z\kern-0.2em Z$}}}}

%
%
\newtheorem{theorem}{Theorem}[section]          
\newtheorem{lemma}[theorem]{Lemma}              
\newtheorem{proposition}[theorem]{Proposition}
\newtheorem{corollary}[theorem]{Corollary}
\newtheorem{definition}[theorem]{Definition}
\newtheorem{conjecture}[theorem]{Conjecture}
\newtheorem{claim}[theorem]{Claim}
\newtheorem{observation}[theorem]{Observation}
\def\proof{\par\noindent{\it Proof.\ }}
\def\reff#1{(\ref{#1})}
%
%
\let\zed=\bbbz 
\let\szed=\bbbz 
\let\IR=\bbbr 
\let\R=\bbbr 
\let\sIR=\bbbr 
\let\IN=\bbbn 
\let\IC=\bbbc 

\def\nl{\medskip\par\noindent}

\def\scrb{{\cal B}}
\def\scrg{{\cal G}}
\def\scrf{{\cal F}}
\def\scrl{{\cal L}}
\def\scrr{{\cal R}}
\def\scrt{{\cal T}}
\def\pfin{{\cal S}}
\def\prob{M_{+1}}
\def\cql{C_{\rm ql}}
\def\bydef{\stackrel{\rm def}{=}}   
\def\qed{\hbox{\hskip 1cm\vrule width6pt height7pt depth1pt \hskip1pt}\bigskip}
\def\remark{\medskip\par\noindent{\bf Remark:}}
\def\remarks{\medskip\par\noindent{\bf Remarks:}}
\def\example{\medskip\par\noindent{\bf Example:}}
\def\examples{\medskip\par\noindent{\bf Examples:}}
\def\nonexamples{\medskip\par\noindent{\bf Non-examples:}}

\newenvironment{scarray}{
          \textfont0=\scriptfont0
          \scriptfont0=\scriptscriptfont0
          \textfont1=\scriptfont1
          \scriptfont1=\scriptscriptfont1
          \textfont2=\scriptfont2
          \scriptfont2=\scriptscriptfont2
          \textfont3=\scriptfont3
          \scriptfont3=\scriptscriptfont3
        \renewcommand{\arraystretch}{0.7}
        \begin{array}{c}}{\end{array}}

\def\wspec{w'_{\rm special}}
\def\mup{\widehat\mu^+}
\def\mupm{\widehat\mu^{+|-_\Lambda}}
\def\pip{\widehat\pi^+}
\def\pipm{\widehat\pi^{+|-_\Lambda}}
\def\ind{{\rm I}}
\def\const{{\rm const}}

\bibliographystyle{plain}


\title{\vspace*{-2.4cm} Pathological Behavior\break
of Renormalization-Group Maps\break
at High Fields\break
and Above the Transition Temperature}

\author{
  \\
  {\normalsize Aernout C. D. van Enter}        \\[-1mm]
  {\normalsize\it Institute for Theoretical Physics}   \\[-1.5mm]
  {\normalsize\it Rijksuniversiteit Groningen}         \\[-1.5mm]
  {\normalsize\it P.O. Box 800}                \\[-1.5mm]
  {\normalsize\it 9747 AG Groningen}           \\[-1.5mm]
  {\normalsize\it THE NETHERLANDS}             \\[-1mm]
  {\normalsize\tt AENTER@TH.RUG.NL}        \\[-1mm]
  {\protect\makebox[5in]{\quad}}  
  \\[-1mm]  \and
  {\normalsize Roberto Fern\'andez}\thanks{Address from October 1,
1994:  FAMAF, Universidad Nacional de C\'ordoba, Ciudad Universitaria, 5000
C\'ordoba, Argentina.  E-mail: {\tt mafcor!fernandez@uunet.uu.net}}\\[-1mm]
  {\normalsize\it Institut de Physique Th\'eorique}                \\[-1.5mm]
  {\normalsize\it Ecole Polytechnique F\'ed\'erale de Lausanne}    \\[-1.5mm]
  {\normalsize\it PHB -- Ecublens}             \\[-1.5mm]
  {\normalsize\it CH--1015 Lausanne}           \\[-1.5mm]
  {\normalsize\it SWITZERLAND}                 \\[-1mm]
  {\normalsize\tt FERNANDEZ@ELDP.EPFL.CH}       \\[-1mm]
  {\protect\makebox[5in]{\quad}}  
  \\[-1mm]  \and
  {\normalsize Roman Koteck\'y}\thanks{On leave from
Center for Theoretical Study, Charles University, Jilsk\'a 1, 110 00
Praha 1, Czech Republic.  E-mail: {\tt KOTECKY@ACI.CVUT.CZ}}
                  \\[-1mm]
  {\normalsize\it CNRS}       \\[-1.5mm]
  {\normalsize\it Luminy -- Case 907}         \\[-1.5mm]
  {\normalsize\it 13288 Marseille Cedex 9}          \\[-1.5mm]
  {\normalsize\it France}                         \\[-1mm]
  {\normalsize\tt KOTECKY@CPT.UNIV-MRS.FR}          \\[-1mm]
  {\protect\makebox[5in]{\quad}}  
  \\[-2mm]
}

\date{\datetitle}

\maketitle
\thispagestyle{empty}

\clearpage

\setcounter{page}{1}

\begin{abstract}
We show that decimation transformations applied to high-$q$ Potts
models result in non-Gibbsian measures even for temperatures higher
than the transition temperature.  We also show that majority
transformations applied to the Ising model in a very strong field at
low temperatures produce non-Gibbsian measures.  This shows that
pathological behavior of renormalization-group transformations is even
more widespread than previous examples already suggested.
\end{abstract}

\tableofcontents

\section{Introduction}

In \cite{vEFS_PRL,vEFS_JSP} it was shown how various
renormalization-group (RG) maps acting on Gibbs measures produce
non-Gibbsian measures.  In physicists' language, this means that a
``renormalized Hamiltonian'' can not be defined.  The examples
presented there were all valid at low temperatures and mostly either
in or close to the coexistence region.  The underlying mechanism
---~pointed out first by Griffiths, Pearce and Israel
 \cite{gripea78,gripea79,isr79}~--- is the fact
that for the constraints imposed by particular choices of block-spin
configurations, the resulting system exhibits a first-order phase
transition.  For this to happen, it was expected that the original
system should be itself at or in the vicinity of a phase transition.
Block-average transformations, however, provided a
counter-example to this belief, in that they lead to
non-Gibbsianness for arbitrary values of the magnetic field (at low
temperatures) \cite{vEFS_JSP}.

Since this work was done, there was a sort of ``damage-control''
movement where various transformations where shown, c.q.\ argued, to
preserve Gibbsianness, or to restore it after sufficiently many
iterations.   These include sufficiently sparse (or sufficiently often
iterated) decimations in nonzero field \cite{maroli93}, possibly
combined with other block-spin transformations \cite{maroli94},
decimated projections on a hyperplane \cite{lorvel94}, and majority
\cite{ken92}, block-average \cite{benmaroli94} and decimation
\cite{vel94} transformations in
the (low-temperature) vicinity of the critical point of the
two-dimensional Ising model.  The case of decimated projections
\cite{lorvel94} has the peculiarity that the Gibbsianness is
restored in a measure-dependent fashion:  the renormalized Hamiltonians
for the ``$+$'' and the ``$-$'' Gibbs states are different, and there
is no renormalized Hamiltonian for nontrivial mixtures of these
states.  On the
other hand, the studies of the 2-$d$ critical Ising model
\cite{ken92,benmaroli94,vel94}, though highly suggestive, are not
conclusive because they involve only (judiciously) selected
block-spin configurations.  Of related interest are the
transformations presented in \cite{hag93.1,hag93.2,hag94} which are
``anti-pathological'' in the sense that they can produce Gibbs
measures out of non-Gibbsian ones.

In this paper we present two new examples of non-Gibbsianness that
show the ubiquity of this phenomenon of lack of a renormalized
Hamiltonian:  1) We show another example of
non-Gibbsianness in the {\em strong-field}\/ region, this time for
majority-rule transformations of the Ising model.
2) For the high-$q$ Potts model we show that the decimated
measure can be non-Gibbsian for a range of temperatures {\em above}\/
the transition temperature.
The first example together with the example of block-averaging
\cite{vEFS_JSP} show that non-Gibbsianness can appear deep within the
region of complete analyticity \cite{dobshl87}, contradicting the
intuition explained in \cite{maroli93,benmaroli94}.
On the other hand, the second example, besides being the first proven
example of a
``high-temperature'' pathology, shows that the condition
of complete analyticity may be violated above the transition
temperature, answering a question posed by Roland Dobrushin.

We mention that Griffiths and Pearce \cite{gripea78,gripea79}, and
also Hasenfratz and Hasenfratz \cite{hashas88}, presented arguments
suggesting the existence of ``peculiarities'' for majority-rule
transformations at some precisely tuned (high) values of the magnetic
field.  Our discussion shows that the situation is even worse than
they expected because in fact the pathologies happen for {\em
arbitrarily large}\/ values of the field.

The present examples, in our opinion, support the point of view that
the non-Gibbsianness of renormalized measures is in some sense
``typical'', and should not be dismissed as exceptional.  On the other
hand, they make even more apparent the need for a systematic study of
the consequences of this non-Gibbsianness on computational schemes
(renormalization-group calculations, image-processing algorithms)
which assume the existence of a renormalized Hamiltonian in the usual
sense (see \cite{sal94} for a pioneer study in this direction).

\section{Basic Set-up}

We consider finite-spin systems in
the lattice $\scrl=\zed^d$, that is a
{\em configuration space}\/ of the
form $\Omega=(\Omega_0)^{\zed^d}$
with the {\em single-spin space}
$\Omega_0$ consisting of a finite set of (integer) numbers.  We consider the
usual
structures:  All subsets of $\Omega_0$ are declared to be open
(discrete topology) and measurable (discrete $\sigma$-algebra), and
the normalized counting measure is chosen as the
{\em a priori}\/ probability
measure on the single-spin space.  The space $\Omega$ is endowed with
the corresponding product structures.   In particular, the product of
normalized counting measures acts as an a-priori probability measure on
$\Omega$  ---~the {\em interaction-free}\/ measure~--- which we denote
$\mu^0$.  We shall use a subscript $\Lambda$ when referring to
analogous objects for a subset $\Lambda\subset\zed^d$: for instance
$\Omega_\Lambda\equiv(\Omega_0)^\Lambda$; if $\sigma\in\Omega$,
$\sigma_\Lambda\equiv(\sigma_x)_{x\in\Lambda}$, etc.  On the other
hand for $\sigma,\omega\in\Omega$ we shall denote
$\sigma_\Lambda\omega$ the configuration equal to $\sigma$ on sites in
$\Lambda$ and to $\omega$ outside.

We point out that, in contrast with the single-spin case, not
all subsets of $\Omega$ are open, nor all functions on $\Omega$
continuous.  In fact, a function $f\colon \Omega\to\R$ is continuous
at $\sigma$ if and only if:
\begin{equation}
   \lim\limits_{\Lambda\nearrow\scrl}\,
   \sup\limits_{\omega\,\colon\, \omega_\Lambda = \sigma_\Lambda}
   \, |f(\sigma) - f(\omega)|  \;=\;   0 \;,
\label{1}
\end{equation}
that is, a change of $\sigma$ in far-away sites has little effect on
the value of $f$.  That is why continuous functions are, in the
present setting, often also called {\em quasilocal}\/ functions.
Here and in the sequel we use
the symbol ``$\nearrow$'' to indicate convergence
in the van Hove sense.  Also, we point out that the symbol ``$|\;|$''
will also be used to indicate the cardinality of a set.

Each spin model is usually defined in terms of an interaction, that
is, a family
$\Phi = (\Phi_A)_{A \subset \zed^d,\;  A\;{\rm finite}}$
of functions $\Phi_A\colon\; \Omega \to \R$ (contribution of the spins
in $A$ to the interaction energy) which are continuous and
depend only on the spins in $A$.  These interactions determine the
finite-volume Hamiltonians
\begin{equation}
H_\Lambda(\sigma_\Lambda|\omega) \;\equiv\;
      \sum_{\begin{scarray}
              {\rm finite}\, A \subset \scrl   \\
              A \cap \Lambda \neq \emptyset
            \end{scarray}
           }
      \Phi_A (\sigma_\Lambda\omega)\;,
\label{ss.2}
\end{equation}
and the Boltzmann-Gibbs weights
\begin{equation}
   \pi_\Lambda(g|\omega) \;=\;
  ({\rm Norm.})^{-1}
    \int g(\sigma_\Lambda\omega) \,
         \exp[-H_\Lambda(\sigma_\Lambda|\omega)]
              \,\mu^0_\Lambda(d\sigma_\Lambda)\;.
\label{ss.3}
\end{equation}
In order not to run into problems with the definition of $H_\Lambda$
and the Boltzmann weights, the usual assumption is that the
interactions are {\em absolutely summable}\/ i.e.\
$\sup_x\sum_{A\ni x} \|\Phi_A\|_\infty <\infty$.

The set of Boltzmann weights $\pi(\,\cdot\,|\,\cdot\,)$ form a regular
system of conditional probabilities in the sense that they satisfy the
``consistence property''
\begin{equation}
\pi_{\tilde\Lambda}(\,\cdot\,|\omega) \;=\;
\int \pi_{\Lambda}(\,\cdot\,|\widetilde\omega) \,
\pi_{\tilde\Lambda}(d\widetilde\omega|\omega)
\label{ss.6}
\end{equation}
for {\em all}\/ configurations $\omega\in\Omega$ and all volumes
$\Lambda \subset \widetilde\Lambda$.  For this reason, they constitute
a system of regular conditional probabilities (for events on finite
volumes
conditioned on the configurations outside).  Moreover, these are
conditional probabilities defined for {\em all}\/ configurations
$\omega$, rather than almost all as is usually the case in probability
theory.  To emphasize this fact, the term {\em specification} has been
coined.

Specifications defined as in \reff{ss.3} are called {\em Gibbsian
specifications}\/, and they model finite-volume equilibrium for the
system in question.  The corresponding infinite-volume equilibrium is
described by the corresponding {\em Gibbs measures}\/, which are those
measures $\mu$
on $\Omega$
whose conditional probabilities are given by the
specification:
\begin{equation}
\mu(\,\cdot\,) \;=\;
\int \pi_{\Lambda}(\,\cdot\,|\omega) \, \mu(d\omega)\;.
\label{ss.7}
\end{equation}
In this case one also says that the measure $\mu$ is {\em
consistent}\/ with the specification $\pi$.  More generally, a
probability measure is {\em Gibbsian}\/ if it is consistent with some
Gibbsian specification.

There is an important necessary condition of Gibbsianness:
Gibbsian specifications are
necessarily
 continuous ---~that is, quasilocal~---
with respect to the boundary conditions.  That is, [c.f.\ \reff{1}],
for each finite $\Lambda\subset\zed^d$,
and any $\sigma\in\Omega$,
\begin{equation}
   \lim\limits_{\Lambda\nearrow\scrl}\,
   \sup\limits_{\omega\,\colon\, \omega_\Lambda = \sigma_\Lambda}
   \, |\pi_\Lambda(\,\cdot\,|\sigma) -
\pi_\Lambda(\,\cdot\,|\omega)|  \;=\;   0
\label{111}
\end{equation}
with the limit understood in the weak sense (i.e.\ it holds,
possibly at different rates, when ``$\,\cdot\,$'' is replaced by any
continuous function depending only on finitely many spins).  A measure
whose conditional probabilities violate this quasilocality requirement
can not be Gibbsian (see \cite{vEFS_JSP} for a more detailed
discussion of this issue).

In particular it is of interest to analyze the Gibbsianness of
renormalized measures.  In its general form, a {\em renormalization
transformation}\/ is a map between probability measures defined by a
probability kernel (see \cite{vEFS_JSP} for the relevant definitions).
In this paper we consider only {\em deterministic real-space
renormalization transformations}\/.  These are defined
in the following
fashion.  One considers a basic ``block'' $B_0$
---~in this paper a cube of linear size $N$~--- and paves $\zed^d$
with
its
translates $\{B_x\,\colon\, x\in N\zed^d\}$ (from now on,
whenever we speak about ``blocks'' we shall mean one of the blocks
of a fixed paving).  For each block one
takes a transformation that associates to each configuration in the
block $B_x$ a spin value representing an ``effective'' block spin.
It is mathematically convenient to think
of this
transformation as going
from $\zed^d$ to $\zed^d$, rather than to a ``thinned'' $\zed^d$,
hence we consider  maps $T_x\,\colon\,\Omega_{B_x}\to\Omega_0$,
defined for each $x\in\zed^d$,
and the map $T\,\colon\,\Omega\to\Omega$ with
$[T(\omega)]_x = T_{Nx}(\omega_{B_{Nx}})$ constructed from it.
Each such map $T$ defines a renormalization transformation on measures
that maps
every
measure $\mu$ on $\Omega$ into a new measure $T\mu$, also
on $\Omega$,
introduced in a natural manner by its action
on any measurable function $g$, namely,
\begin{equation}
\int g(\omega') \,T\mu(d\omega') \;=\; \int g\bigl(T(\omega)\bigr)
\mu(d\omega)\;.
\label{ss.10}
\end{equation}
(As customary, we shall try to use primed variables for the
renormalized objects.)  The two transformations of interest here are
odd-block majority-rule transformations for the Ising model
($\sigma_x=+1,-1$):
\begin{equation}
T_x\sigma_{B_x} \;=\; {\rm sgn}\,\left(\sum_{x\in B_x}
\sigma_x\right)\;,
\label{ss.11}
\end{equation}
and decimation for the Potts model
\begin{equation}
T_x(\sigma_{B_x}) \;=\; \sigma_x\;.
\label{ss.12}
\end{equation}

\section{Non-Gibbsianness for Majority-Rule Maps\hspace*{\fill}\break
of Ising Models at High Magnetic Field} \label{smaj}

We consider the Ising model in $\zed^d$, that is spins
$\sigma_x\in\{-1,1\}$ with interaction
\begin{equation}
   \Phi_A(\sigma)   \;=\;
   \cases{  -h \sigma_x              & if $A = \{x\}$   \cr
            -J \sigma_x \sigma_y  & if $A = \{x,y\}$ \hbox{ with }
x,y \hbox{ nearest neighbors}\cr
            0                          & otherwise $\;$,     \cr
         }
\label{4}
\end{equation}
with $J>0$.  The result is the following:

\begin{theorem}\label{tmaj}
Consider the majority-rule transformation $T_L$ acting on
blocks of linear size $2L+1$, $L\ge2$.  Let $\mu_{\beta, h}$
denote the unique Gibbs measure for the Ising model at inverse
temperature $\beta$ and magnetic field $h>0$.  Then there exists a
$\beta_L$ such that for $\beta>\beta_L$ and $|h|>J/L$ the measure
$T_L\mu_{\beta,h}$ is not consistent with any quasilocal
specification; in particular, it is not
a Gibbs measure for any
uniformly convergent interaction.
\end{theorem}

For the proof we essentially follow the scheme of \cite[Section
4.2]{vEFS_JSP}:  We determine a suitable special configuration
$\wspec$ yielding a constrained system with several phases.
Let us, for concreteness, consider $h>0$.  In this case we choose
$\wspec$ equal to the all-``$-$'' configuration, so as to have a
constraint acting against the magnetic field.   We have to prove two
things:
\begin{claim}\label{claim1}
The resulting constrained system of internal spins has more than one
phase.
\end{claim}

\begin{claim}\label{claim2}
The different phases
of the constrained system can be selected by imposing
suitable {\em block-spin}\/ boundary conditions, over a ring-like
region of finite width
(i.e. by replacing, for this boundary
region, the above constraint stemming from $\wspec$
 by a different suitably chosen constraint).
\end{claim}

Together these claims imply that by changing block spins arbitrarily
far away, one changes the phase of the internal spins, which in turns
changes the value of block-spin averages close to the origin.  For
instance it modifies the (average) value of the block-spin at the
origin and that of one
of its nearest-neighbors (when these spins are ``unfixed'';
this part of the argument is almost identical to the corresponding
argument for block-averaging transformations; see Step 3 in \cite[pp.\
1008-1009]{vEFS_JSP}.)   This modification takes place despite the fact
that the intermediate block spins are fixed in the configuration
$\wspec$.  This means that the {\em direct}\/ influence of far away
block spins does not decrease with the distance, hence the renormalized
measure can not be Gibbsian.

We emphasize that only block spins on an annulus of {\em finite
width}\/ are invoked in Claim \ref{claim2}; the block-spin
configurations can be arbitrarily chosen outside it.  This implies
that there is an
``essential'' jump in averages of renormalized observables, in which
the extremal values of it can be reached via sequences chosen from
``large'' (non-zero-measure) sets of boundary configurations, obtained
by modifying $\wspec$ arbitrarily far away.
Mathematically, we are proving that some
conditional probabilities of $T_L\mu_{\beta, h}$ are {\em
essentially}\/ discontinuous at $\wspec$:  They exhibit a jump that
can not be removed by redefining them on a set of $\mu_{\beta,
h}$-measure zero around $\wspec$.  Hence, no other realization of such
conditional probabilities will be free of this discontinuity.  Of
course, one may attempt to do without $\wspec$; after all conditional
probabilities need to be defined only $T_L\mu_{\beta, h}$-almost
everywhere.  This is a more involved issue about which we shall briefly
comment in Section \ref{sconc}.   The finiteness of the annulus in
Claim \ref{claim2} is needed for a second reason:  A priori we only
know that the conditional probabilities of $T_L\mu_{\beta, h}$ are
{\em some}\/ Gibbs states of the constrained system of internal spins
[see the discussion of Step 0 (esp.\ pages 987--990) in
\cite{vEFS_JSP}],
but we do not know which ones.  Therefore, the statements have to be
proved for {\em all}\/ possible such Gibbs states, which is equivalent
\cite[Theorem 7.12]{geo88} to proving them for {\em arbitrary boundary
conditions}\/ (see \cite[p.\ 991]{vEFS_JSP} for a more
complete discussion of these issues).
\medskip

We discuss the proof of the claims above only in the particular case
of $d=2$ and $L=2$  ($5\times 5$--blocks).  The other cases are
analogous, but they require a more complicated accounting of ground
states that would obscure the argument.
\bigskip

\subsection{Proof of Claim \protect\ref{claim1}}

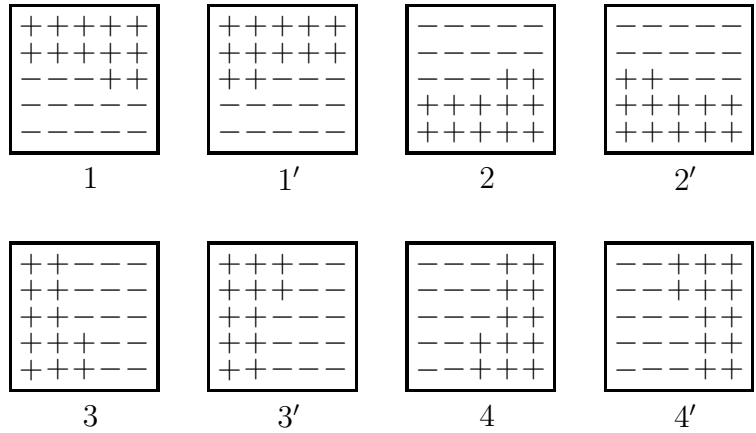
\begin{figure}[p]
\begin{center}
\begin{picture}(285,165)(-142.5,-82.5)
%
%
\put(-142.5,7.5){
\begin{picture}(60,75)(-30,-45)
\thicklines
\put(-30,-30){\framebox(55,55){
\begin{picture}(50,50)(-25,-25)
\multiput(-20,20)(10,0){5}{\makebox(0,0){$+$}}
\multiput(-20,10)(10,0){5}{\makebox(0,0){$+$}}
\multiput(-20,0)(10,0){3}{\makebox(0,0){$-$}}
\multiput(10,0)(10,0){2}{\makebox(0,0){$+$}}
\multiput(-20,-10)(10,0){5}{\makebox(0,0){$-$}}
\multiput(-20,-20)(10,0){5}{\makebox(0,0){$-$}}
\end{picture}
}}
\put(0,-40){\makebox(0,0){$1$}}
\end{picture}
}
\put(-67.5,7.5){
\begin{picture}(60,75)(-30,-45)
\thicklines
\put(-30,-30){\framebox(55,55){
\begin{picture}(50,50)(-25,-25)
\multiput(-20,20)(10,0){5}{\makebox(0,0){$+$}}
\multiput(-20,10)(10,0){5}{\makebox(0,0){$+$}}
\multiput(-20,0)(10,0){2}{\makebox(0,0){$+$}}
\multiput(0,0)(10,0){3}{\makebox(0,0){$-$}}
\multiput(-20,-10)(10,0){5}{\makebox(0,0){$-$}}
\multiput(-20,-20)(10,0){5}{\makebox(0,0){$-$}}
\end{picture}
}}
\put(0,-40){\makebox(0,0){$1'$}}
\end{picture}
}
\put(7.5,7.5){
\begin{picture}(60,75)(-30,-45)
\thicklines
\put(-30,-30){\framebox(55,55){
\begin{picture}(50,50)(-25,-25)
\multiput(-20,20)(10,0){5}{\makebox(0,0){$-$}}
\multiput(-20,10)(10,0){5}{\makebox(0,0){$-$}}
\multiput(-20,0)(10,0){3}{\makebox(0,0){$-$}}
\multiput(10,0)(10,0){2}{\makebox(0,0){$+$}}
\multiput(-20,-10)(10,0){5}{\makebox(0,0){$+$}}
\multiput(-20,-20)(10,0){5}{\makebox(0,0){$+$}}
\end{picture}
}}
\put(0,-40){\makebox(0,0){$2$}}
\end{picture}
}
\put(82.5,7.5){
\begin{picture}(60,75)(-30,-45)
\thicklines
\put(-30,-30){\framebox(55,55){
\begin{picture}(50,50)(-25,-25)
\multiput(-20,20)(10,0){5}{\makebox(0,0){$-$}}
\multiput(-20,10)(10,0){5}{\makebox(0,0){$-$}}
\multiput(-20,0)(10,0){2}{\makebox(0,0){$+$}}
\multiput(0,0)(10,0){3}{\makebox(0,0){$-$}}
\multiput(-20,-10)(10,0){5}{\makebox(0,0){$+$}}
\multiput(-20,-20)(10,0){5}{\makebox(0,0){$+$}}
\end{picture}
}}
\put(0,-40){\makebox(0,0){$2'$}}
\end{picture}
}
%
%
\put(-142.5,-82.5){
\begin{picture}(60,75)(-30,-45)
\thicklines
\put(-30,-30){\framebox(55,55){
\begin{picture}(50,50)(-25,-25)
\multiput(-20,-20)(0,10){5}{\makebox(0,0){$+$}}
\multiput(-10,-20)(0,10){5}{\makebox(0,0){$+$}}
\multiput(0,-20)(0,10){2}{\makebox(0,0){$+$}}
\multiput(0,0)(0,10){3}{\makebox(0,0){$-$}}
\multiput(10,-20)(0,10){5}{\makebox(0,0){$-$}}
\multiput(20,-20)(0,10){5}{\makebox(0,0){$-$}}
\end{picture}
}}
\put(0,-40){\makebox(0,0){$3$}}
\end{picture}
}
\put(-67.5,-82.5){
\begin{picture}(60,75)(-30,-45)
\thicklines
\put(-30,-30){\framebox(55,55){
\begin{picture}(50,50)(-25,-25)
\multiput(-20,-20)(0,10){5}{\makebox(0,0){$+$}}
\multiput(-10,-20)(0,10){5}{\makebox(0,0){$+$}}
\multiput(0,-20)(0,10){3}{\makebox(0,0){$-$}}
\multiput(0,10)(0,10){2}{\makebox(0,0){$+$}}
\multiput(10,-20)(0,10){5}{\makebox(0,0){$-$}}
\multiput(20,-20)(0,10){5}{\makebox(0,0){$-$}}
\end{picture}
}}
\put(0,-40){\makebox(0,0){$3'$}}
\end{picture}
}
\put(7.5,-82.5){
\begin{picture}(60,75)(-30,-45)
\thicklines
\put(-30,-30){\framebox(55,55){
\begin{picture}(50,50)(-25,-25)
\multiput(-20,-20)(0,10){5}{\makebox(0,0){$-$}}
\multiput(-10,-20)(0,10){5}{\makebox(0,0){$-$}}
\multiput(0,-20)(0,10){2}{\makebox(0,0){$+$}}
\multiput(0,0)(0,10){3}{\makebox(0,0){$-$}}
\multiput(10,-20)(0,10){5}{\makebox(0,0){$+$}}
\multiput(20,-20)(0,10){5}{\makebox(0,0){$+$}}
\end{picture}
}}
\put(0,-40){\makebox(0,0){$4$}}
\end{picture}
}
\put(82.5,-82.5){
\begin{picture}(60,75)(-30,-45)
\thicklines
\put(-30,-30){\framebox(55,55){
\begin{picture}(50,50)(-25,-25)
\multiput(-20,-20)(0,10){5}{\makebox(0,0){$-$}}
\multiput(-10,-20)(0,10){5}{\makebox(0,0){$-$}}
\multiput(0,-20)(0,10){3}{\makebox(0,0){$-$}}
\multiput(0,10)(0,10){2}{\makebox(0,0){$+$}}
\multiput(10,-20)(0,10){5}{\makebox(0,0){$+$}}
\multiput(20,-20)(0,10){5}{\makebox(0,0){$+$}}
\end{picture}
}}
\put(0,-40){\makebox(0,0){$4'$}}
\end{picture}
}
\end{picture}
\end{center}
\caption{Configurations minimizing the energy within a $5\times
5$--block for the Ising model with negative block magnetization in the
regime $h>J$.}
\label{f.1}
\end{figure}

\begin{figure}
\begin{center}
\begin{picture}(210,245)(-105,-122.5)
%
%
\put(-105,12.5){
\begin{picture}(85,110)(-42.5,-72.5)
\thicklines
\multiput(-37.5,37.5)(30,0){3}{\makebox(0,0){$1'$}}
\multiput(-22.5,37.5)(30,0){3}{\makebox(0,0){$1$}}
\multiput(-37.5,22.5)(30,0){3}{\makebox(0,0){$2$}}
\multiput(-22.5,22.5)(30,0){3}{\makebox(0,0){$2'$}}
\multiput(-37.5,7.5)(30,0){3}{\makebox(0,0){$1$}}
\multiput(-22.5,7.5)(30,0){3}{\makebox(0,0){$1'$}}
\multiput(-37.5,-7.5)(30,0){3}{\makebox(0,0){$2'$}}
\multiput(-22.5,-7.5)(30,0){3}{\makebox(0,0){$2$}}
\multiput(-37.5,-22.5)(30,0){3}{\makebox(0,0){$1'$}}
\multiput(-22.5,-22.5)(30,0){3}{\makebox(0,0){$1$}}
\multiput(-37.5,-37.5)(30,0){3}{\makebox(0,0){$2'$}}
\multiput(-22.5,-37.5)(30,0){3}{\makebox(0,0){$2$}}
\put(0,-62.5){\makebox(0,0){I}}
\end{picture}
}
\put(20,12.5){
\begin{picture}(85,110)(-42.5,-72.5)
\thicklines
\multiput(-37.5,37.5)(30,0){3}{\makebox(0,0){$2'$}}
\multiput(-22.5,37.5)(30,0){3}{\makebox(0,0){$2$}}
\multiput(-37.5,22.5)(30,0){3}{\makebox(0,0){$1'$}}
\multiput(-22.5,22.5)(30,0){3}{\makebox(0,0){$1$}}
\multiput(-37.5,7.5)(30,0){3}{\makebox(0,0){$2$}}
\multiput(-22.5,7.5)(30,0){3}{\makebox(0,0){$2'$}}
\multiput(-37.5,-7.5)(30,0){3}{\makebox(0,0){$1$}}
\multiput(-22.5,-7.5)(30,0){3}{\makebox(0,0){$1'$}}
\multiput(-37.5,-22.5)(30,0){3}{\makebox(0,0){$2'$}}
\multiput(-22.5,-22.5)(30,0){3}{\makebox(0,0){$2$}}
\multiput(-37.5,-37.5)(30,0){3}{\makebox(0,0){$1$}}
\multiput(-22.5,-37.5)(30,0){3}{\makebox(0,0){$1'$}}
\put(0,-62.5){\makebox(0,0){II}}
\end{picture}
}
%
%
%
\put(-105,-122.5){
\begin{picture}(85,110)(-42.5,-72.5)
\thicklines
\multiput(-37.5,-37.5)(0,30){3}{\makebox(0,0){$3$}}
\multiput(-37.5,-22.5)(0,30){3}{\makebox(0,0){$3'$}}
\multiput(-22.5,-37.5)(0,30){3}{\makebox(0,0){$4'$}}
\multiput(-22.5,-22.5)(0,30){3}{\makebox(0,0){$4$}}
\multiput(-7.5,-37.5)(0,30){3}{\makebox(0,0){$3$}}
\multiput(-7.5,-22.5)(0,30){3}{\makebox(0,0){$3'$}}
\multiput(7.5,-37.5)(0,30){3}{\makebox(0,0){$4'$}}
\multiput(7.5,-22.5)(0,30){3}{\makebox(0,0){$4$}}
\multiput(22.5,-37.5)(0,30){3}{\makebox(0,0){$3'$}}
\multiput(22.5,-22.5)(0,30){3}{\makebox(0,0){$3$}}
\multiput(37.5,-37.5)(0,30){3}{\makebox(0,0){$4$}}
\multiput(37.5,-22.5)(0,30){3}{\makebox(0,0){$4'$}}
\put(0,-62.5){\makebox(0,0){III}}
\end{picture}
}
\put(20,-122.5){
\begin{picture}(85,110)(-42.5,-72.5)
\thicklines
\multiput(-37.5,-37.5)(0,30){3}{\makebox(0,0){$4$}}
\multiput(-37.5,-22.5)(0,30){3}{\makebox(0,0){$4'$}}
\multiput(-22.5,-37.5)(0,30){3}{\makebox(0,0){$3'$}}
\multiput(-22.5,-22.5)(0,30){3}{\makebox(0,0){$3$}}
\multiput(-7.5,-37.5)(0,30){3}{\makebox(0,0){$4'$}}
\multiput(-7.5,-22.5)(0,30){3}{\makebox(0,0){$4$}}
\multiput(7.5,-37.5)(0,30){3}{\makebox(0,0){$3$}}
\multiput(7.5,-22.5)(0,30){3}{\makebox(0,0){$3'$}}
\multiput(22.5,-37.5)(0,30){3}{\makebox(0,0){$4$}}
\multiput(22.5,-22.5)(0,30){3}{\makebox(0,0){$4'$}}
\multiput(37.5,-37.5)(0,30){3}{\makebox(0,0){$3'$}}
\multiput(37.5,-22.5)(0,30){3}{\makebox(0,0){$3$}}
\put(0,-62.5){\makebox(0,0){IV}}
\end{picture}
}
\end{picture}
\end{center}
\caption{Classes of ground states for the Ising model with negative
block magnetization ($5\times 5$--block, $h>J$).  Within each strip
the primed blocks can either be at odd or at even positions,
independently of the configuration in other strips.}
\label{f.2}
\end{figure}
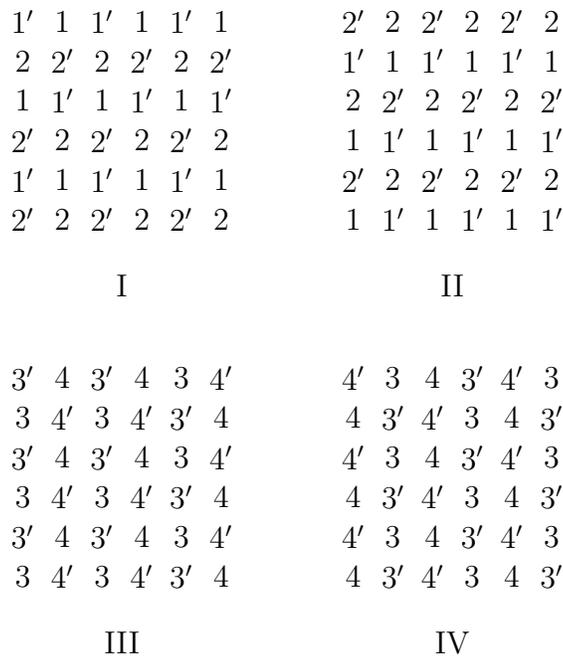

We start by analyzing the ground-state configurations of the
constrained system.  These configurations must satisfy the
constraint of keeping each block with a majority of ``$-$'', while
maximizing the number of spins parallel to the field and minimizing
the number of ``$+$''-``$-$'' pairs (broken bonds).
This clearly yields, inside  $5\times 5$ blocks and for $h>J$,
the 8 ground state configurations shown in  Figure
\ref{f.1}.
 Any overall ground state configuration combines
such blocks without any interruption. It is easy to convince
oneself that there is an infinite number of such ground state
configurations and that this set splits into four classes
consisting of configurations with either horizontal or vertical
alternating strips as depicted in Figure
\ref{f.2}.  Within each strip a primed block always neighbors an
unprimed one and one has the freedom to start, in each strip
independently of the other strips, with the primed or unprimed
one.
This yields  two possible  arrangements [mapped one into another
by  a shift  by one (block) lattice spacing]
for each strip and leads to the  degeneracy of the order
$2^{\rm number\; of\; strips}$ of each of these classes of ground
state configurations.

We assert that each class of ground
state configurations
 gives rise to a different
low-temperature Gibbs measure.  In such measures
only the identity of the class is kept --- the periodic
long-range order between primed and unprimed blocks
present in particular ground configurations is not conserved at
nonvanishing temperatures as it is, effectively,
a one-dimensional order.  The proof of this assertion, from which
Claim \ref{claim1} follows, can be done in (at least) two different
ways.

 The first one is
to use chessboard estimates in the form presented in
Theorem 18.25 of \cite{geo88}.  Indeed, by considering each block as
a single-spin space with as many values as block configurations
satisfying the constraint of having a majority ``$-$'', we can map
our constrained system into an unconstrained one with
$|\Omega_0|=2^{24}$ and
with a certain
one- and two-body nearest-neighbor
interaction.  This system is clearly reflection-positive and the
four classes of Figure \ref{f.2} are
just
the classes $G_1,\ldots,G_4$ of
the above mentioned theorem.

One can also prove the existence of four low-temperature Gibbs states
with the help of
the generalization of Pirogov-Sinai theory due to
Bricmont, Kuroda and Lebowitz (BKL) \cite{brikurleb85}.  Let us
briefly review BKL theory, as we also apply it later for the
example of the Potts model.  The central objects of the theory are
the {\em restricted ensembles}\/ which are families or classes of
configurations
that
play a r\^ole analogous to that
of
the ground states
in the standard Pirogov-Sinai theory.  In BKL
version, the restricted ensembles have a product structure:  they
are characterized by their configurations on
an elementary
cube $C_0$.  More precisely,
$\Omega_{C_0}$ can be partitioned,
\begin{equation}
\Omega_{C_0} \;=\; \left[\bigcup_{a=1}^r \Omega_0^a \right]
\,\cup\, \overline\Omega_0\;,
\label{set.1}
\end{equation}
with
each $\Omega_0^a$
associated to a restricted ensemble
and $\overline\Omega_0$
containing what is left.
By paving the lattice with
translates $C_x$ of $C_0$ with $x\in L\zed^d$,
where $L$ is the linear size of $C_0$, one defines the
translated cube-configurations $\Omega_x^a$.  The $a$-th
restricted ensemble is formed by configurations whose
restriction to each $C_x$ is of the type $\Omega_x^a$:
\begin{equation}
\Omega^a \;=\; \Bigl\{ \sigma\in\Omega \,\colon\, \sigma_{C_x}\in
\Omega_x^a \,
\hbox{ for all }
x\in L\zed^d\Bigr \}\;.
\label{set.2}
\end{equation}
For each restricted ensemble one considers the corresponding
restricted partition functions
in finite volumes $\Lambda$,
\begin{equation}
Z_R(\Lambda,\omega^a) \;=\; \sum_{\sigma_\Lambda\in\Omega^a_\Lambda}
e^{- H_\Lambda(\sigma_\Lambda|\omega^a)}\;,
\label{set.3}
\end{equation}
with boundary conditions $\omega^a\in\Omega^a$.

To apply BKL
theory, several hypotheses must be satisfied
(hypotheses (A1)--(A5) in \cite{brikurleb85}).  First, there is
the
{\em diluteness hypothesis}\/,
which basically means that the restricted partition functions must
admit a polymer expansion from which a convergent cluster
(high-temperature, Mayer) expansion follows.
The diluteness hypothesis implies, in particular, that the
restricted free energies
\begin{equation}
f^a \;\equiv\;
-\lim_{\Lambda\nearrow\zed^d} \frac{1}{|\Lambda|}
\,\log Z_R(\Lambda,\omega^a)
\label{set.4}
\end{equation}
exist and are independent of the choice of $\omega^a\in\Omega^a$.
Second, one assumes a restricted-ensemble
{\em Peierls condition}\/, i.e.\ that the free-energy cost
of placing a droplet of configurations of one of the restricted
ensembles inside a sea corresponding to another restricted ensemble be
proportional to the surface of the droplet.  An important r\^ole is
played by the value, $\tau$, of the constant of proportionality.
Third, the system must
exhibit {\em free-energy}\/ degeneracy among the restricted ensembles:
\begin{equation}
f^a \;=\; f^b \quad 1\le a,b\le r\;.
\label{set.5}
\end{equation}
If restricted ensembles are formed by exactly one configuration, then
the restricted free energies are just energy densities; in that case
\reff{set.5} is the usual degeneracy condition of ground states.
BKL also assumes the existence of
$r-1$ sufficiently smooth perturbations of the interaction, modulated
by parameters $\underline\mu=(\mu_1,\ldots,\mu_{r-1})$, which are
degeneracy-lifting in the sense that the perturbed restricted free
energies $f^a_{\underline\mu}$ produce a phase diagram that obeys the
Gibbs phase rule.  More explicitly, the manifolds in
$\underline\mu$-space defined by inequalities of the form
$f^{a_1}_{\underline\mu}=\cdots=f^{a_k}_{\underline\mu}<
f^{a_{k+1}}_{\underline\mu},\cdots,f^{a_r}_{\underline\mu}$
(``manifolds of $k$-phase coexistence''), can be
homeomorphically mapped, for $\underline\mu$ small enough, onto
an $(r-k)$-dimensional
hypersurfaces of the boundary of the positive
$r$-octant in $\R^r$.  In particular $\underline\mu=\underline 0$ is
the only value for which all the restricted free energies coincide.

Under these hypotheses, the conclusion of BKL theory is that {\em for
$\tau$ large enough}\/ the actual phase diagram of the system is only
a small perturbation of that one drawn with the restricted free
energies.  In particular there is a value $\underline\mu_0$ of the
parameters for which all
$r$ phases associated to the respective
restricted ensembles coexist.  Moreover, this coexistence happens for
\begin{equation}
\|\underline\mu_0\|_\infty \;<\; \const\, e^{-\tau}\;,
\label{set.6}
\end{equation}
that is, the distance between the true maximal-coexistence point and
the one determined via the
 restricted-ensembles by \reff{set.5} tends
exponentially to zero with the Peierls constant.
The typical configurations of the different Gibbs states are formed by
an infinite sea of spins configured as in the corresponding restricted
ensemble, with small bubbles here and there configured as in the other
ensembles.
\bigskip

It is clear how to apply BKL theory for the case of interest here:
The restricted ensembles are the four classes
$\Omega^I,\ldots,\Omega^{IV}$ obtained from the corresponding
configurations of Figure \ref{f.2} by allowing a free assignment of
the primes.
Notice that we extend the original classes of ground
configurations by ignoring the (fake) one-dimensional
primed-unprimed order.
In spite of the fact that restricted excitations are included, the
classes keep their identity and, in particular, the Peierls
condition may be verified.
 For each restricted ensemble, the
restricted partition function is (can be put in correspondence
with) a product of partition functions for one-dimensional
antiferromagnetic Ising models with nearest neighbor coupling $-J$
(the ``primes'' of different lines do not interact, and two
consecutive primes or two consecutive non-primes along a line cost
an energy $J$).  The partition functions for one-dimensional
finite-range systems have all the diluteness properties in the
world, and the four classes have the same restricted free energy
density.
Explicitly, one can easily verify the diluteness
hypothesis in the alternative formulation from Section 4
of \cite{brikurleb85}, that is by exhibiting an exponential decay of
truncated correlations.

To verify the Peierls condition, one has to evaluate the
ratio
\begin{equation}
Q(\Gamma| \Lambda, \omega^a)=
\frac{Z(\Gamma| \Lambda, \omega^a)}{Z_R(\Lambda, \omega^a)}
\label{x}
\end{equation}
with $Z(\Gamma| \Lambda, \omega^a)$ denoting the
partition functions obtained by summing over all
configurations in $\Lambda$ having only one contour
$\Gamma$ (the union of blocks that differ from the
minimizing ones shown on Figure \ref{f.1} equals $\Gamma$).
Using the above mentioned effective equivalence of
the restricted ensemble with uncoupled
one-dimensional Ising models, we evaluate (up to boundary
terms) the restricted partition function  $Z_R(\Lambda,
\omega^a)$ by $(1+ e^{-\beta J})^{|\Lambda|}$. Noticing
that every block in $\Gamma$ is disfavored by at least
the factor $e^{-\beta J}$, we get the Peierls condition
with   the Peierls constant being  at least
$\tau\ge\beta J$.
As symmetry-breaking perturbations we can take
fields selecting one or the other of the classes.  BKL theory
implies, therefore, that for low enough temperature there is a set
of values for the fields (not exceeding $e^{-\beta J}$) at which
four Gibbs state coexist which are supported on configurations
that, except for small fluctuations, look like those of the
corresponding restricted ensemble. Symmetry considerations imply
that these coexistence point
actually
occurs when all the perturbing fields
vanish.

This argument proves Claim \ref{claim1}, and constitutes the rigorous
version of the stated breaking of the long-range order between primed
and unprimed blocks.

\subsection{Proof of Claim \protect\ref{claim2}}

We start by noticing that if
volumes $\Lambda$ as in Figure \ref{f.3} had internal-spin boundary
configurations as in part (a) of
the figure [resp.\ part (b)], then the limit $\Lambda\nearrow
\zed^2$ would select the Gibbs measure corresponding to the class
labeled I
[resp. II]
in Figure \ref{f.2}.
\begin{figure}
\begin{center}
\begin{picture}(350,485)(-175,-215)
%
%
\put(-175,15){
\begin{picture}(350,255)(-175,-137.5)
\put(-150,-62.5){\line(1,0){300}}
\multiput(-150,-62.5)(25,25){5}{\line(0,1){25}}
\multiput(-150,-37.5)(25,25){4}{\line(1,0){25}}
\multiput(150,-62.5)(-25,25){5}{\line(0,1){25}}
\multiput(150,-37.5)(-25,25){4}{\line(-1,0){25}}
\put(-50,62.5){\line(1,0){100}}
\multiput(-137.5,-75)(25,0){12}{\makebox(0,0){$25^+$}}
\multiput(-37.5,75)(25,0){4}{\makebox(0,0){$25^+$}}
\multiput(-162.5,-50)(25,25){5}{\makebox(0,0){$25^+$}}
\multiput(162.5,-50)(-25,25){5}{\makebox(0,0){$25^+$}}
\put(-150,-92.5){\vector(1,0){300}}
\put(150,-92.5){\vector(-1,0){300}}
\put(0,-102.5){\makebox(0,0){$3N$}}
\put(-50,92.5){\vector(1,0){100}}
\put(50,92.5){\vector(-1,0){100}}
\put(0,102.5){\makebox(0,0){$N$}}
\put(0,-130){\makebox(0,0){(a)}}
\end{picture}
}
%
%
%
\put(-175,-215){
\begin{picture}(350,200)(-175,-110)
\put(-150,-62.5){\line(1,0){300}}
\multiput(-150,-62.5)(25,25){5}{\line(0,1){25}}
\multiput(-150,-37.5)(25,25){4}{\line(1,0){25}}
\multiput(150,-62.5)(-25,25){5}{\line(0,1){25}}
\multiput(150,-37.5)(-25,25){4}{\line(-1,0){25}}
\put(-50,62.5){\line(1,0){100}}
\multiput(-137.5,-75)(25,0){12}{\makebox(0,0){$25^-$}}
\multiput(-37.5,75)(25,0){4}{\makebox(0,0){$25^-$}}
\multiput(-162.5,-50)(25,25){5}{\makebox(0,0){$25^-$}}
\multiput(162.5,-50)(-25,25){5}{\makebox(0,0){$25^-$}}
\put(0,-100){\makebox(0,0){(b)}}
\end{picture}
}
\end{picture}
\end{center}
\caption{Internal-spin configurations that would select the Gibbs
measure corresponding to ground states (a) of class I (Figure \protect
\ref{f.2}), (b) of class II.
 The symbols $25^+$ [resp.\ $25^-$]
indicate that
the corresponding block
is composed entirely of ``pluses'' (25 of them)   [resp.\
of 25 ``minuses''].}
\label{f.3}
\end{figure}
This can be seen through a small adaptation of
the usual Peierls argument:  the left and right diagonals are
``neutral'' in that they do not favor any of the ground states,
while the top and bottom favor class I over II in case (a), and
conversely in case (b).
Similarly chosen rotated volumes select classes III and IV.

\begin{figure}
\begin{center}
\begin{picture}(400,450)(-200,-225)
%
%
\put(-175,40){
\begin{picture}(400,205)(-200,-112.5)
\put(-150,-62.5){\line(1,0){300}}
\multiput(-150,-62.5)(25,25){5}{\line(0,1){25}}
\multiput(-150,-37.5)(25,25){4}{\line(1,0){25}}
\multiput(150,-62.5)(-25,25){5}{\line(0,1){25}}
\multiput(150,-37.5)(-25,25){4}{\line(-1,0){25}}
\put(-50,62.5){\line(1,0){100}}
%
\multiput(-187.5,-100)(25,0){16}{\makebox(0,0){$+$}}
\multiput(-187.5,-75)(25,0){16}{\makebox(0,0){$+$}}
\multiput(-187.5,-50)(25,0){2}{\makebox(0,0){$+$}}
\multiput(187.5,-50)(-25,0){2}{\makebox(0,0){$+$}}
\multiput(-187.5,-25)(25,0){3}{\makebox(0,0){$+$}}
\multiput(187.5,-25)(-25,0){3}{\makebox(0,0){$+$}}
\multiput(-187.5,0)(25,0){4}{\makebox(0,0){$+$}}
\multiput(187.5,0)(-25,0){4}{\makebox(0,0){$+$}}
\multiput(-187.5,25)(25,0){5}{\makebox(0,0){$+$}}
\multiput(187.5,25)(-25,0){5}{\makebox(0,0){$+$}}
\multiput(-187.5,50)(25,0){6}{\makebox(0,0){$+$}}
\multiput(187.5,50)(-25,0){6}{\makebox(0,0){$+$}}
\multiput(-187.5,75)(25,0){16}{\makebox(0,0){$+$}}
\multiput(-187.5,100)(25,0){16}{\makebox(0,0){$+$}}
%
%
\put(0,-120){\makebox(0,0){(a)}}
\end{picture}
}
%
%
%
\put(-175,-225){
\begin{picture}(400,205)(-200,-112.5)
\put(-150,-62.5){\line(1,0){300}}
\multiput(-150,-62.5)(25,25){5}{\line(0,1){25}}
\multiput(-150,-37.5)(25,25){4}{\line(1,0){25}}
\multiput(150,-62.5)(-25,25){5}{\line(0,1){25}}
\multiput(150,-37.5)(-25,25){4}{\line(-1,0){25}}
\put(-50,62.5){\line(1,0){100}}
%
\multiput(-187.5,-100)(25,0){16}{\makebox(0,0){$+$}}
\multiput(-187.5,-75)(25,0){2}{\makebox(0,0){$+$}}
\multiput(-137.5,-75)(25,0){12}{\makebox(0,0){$-$}}
\multiput(187.5,-75)(-25,0){2}{\makebox(0,0){$+$}}
\multiput(-187.5,-50)(25,0){2}{\makebox(0,0){$+$}}
\multiput(187.5,-50)(-25,0){2}{\makebox(0,0){$+$}}
\multiput(-187.5,-25)(25,0){3}{\makebox(0,0){$+$}}
\multiput(187.5,-25)(-25,0){3}{\makebox(0,0){$+$}}
\multiput(-187.5,0)(25,0){4}{\makebox(0,0){$+$}}
\multiput(187.5,0)(-25,0){4}{\makebox(0,0){$+$}}
\multiput(-187.5,25)(25,0){5}{\makebox(0,0){$+$}}
\multiput(187.5,25)(-25,0){5}{\makebox(0,0){$+$}}
\multiput(-187.5,50)(25,0){6}{\makebox(0,0){$+$}}
\multiput(187.5,50)(-25,0){6}{\makebox(0,0){$+$}}
\multiput(-187.5,75)(25,0){6}{\makebox(0,0){$+$}}
\multiput(-37.5,75)(25,0){4}{\makebox(0,0){$-$}}
\multiput(187.5,75)(-25,0){6}{\makebox(0,0){$+$}}
\multiput(-187.5,100)(25,0){16}{\makebox(0,0){$+$}}
%
%
\put(0,-120){\makebox(0,0){(b)}}
\end{picture}
}
\end{picture}
\end{center}
\caption{Block-spin configurations that yield, with high probability,
the internal-spin configurations of
Figure \protect \ref{f.3}.}
\label{f.4}
\end{figure}

However, we are allowed  to impose
only
{\em block-spin}\/
configurations, which determine the internal spins only in a
probabilistic sense.  We have to prove that there exist some
block-spin configurations which, when imposed on some annulus of {\em
finite}\/ radius
around $\Lambda$, produce
{\em with high probability}\/ the internal-spin configurations of
Figure \ref{f.3}.  As the reader may suspect, such a configuration
will be the all-``$+$'' block-spin configuration for case (a) [Figure
\ref{f.4} (a)].  For case (b) we shall consider the configuration of
Figure \ref{f.4} (b).  Let us discuss the former case; the latter is
just a
slightly modified version of it.  The argument is basically
a combination of Steps 2.1--2.4 of \cite{vEFS_JSP} (cf.\ p.\
1005 there), and well-known probabilistic Peierls arguments
(see for instance \cite[Section 2]{chachafro85}).

The precise statements require further notation.
 For a block $B$, denote
\begin{equation}
N_+(B) \;=\; \hbox{number of ``$+$'' spins in $B$}.
\label{f.4.1}
\end{equation}
For any family $\gamma$ of $5\times 5$-blocks we use
$|\gamma|$ to denote the number of blocks in $\gamma$ (for
a given configuration) and  take
\begin{equation}
\scrb(\gamma) \;=\; \Bigl\{ {\rm blocks}\; B\in\gamma \,\colon\,
N_+(B) < 25 \Bigr\}\;,
\label{5}
\end{equation}
the set of blocks of $\gamma$ with ``bad'' internal-spin
configurations.  For volumes $V$ formed by a union of non-overlapping
blocks  we consider the probability measures
$\pip_V(\,\cdot\,|\sigma)$,
obtained from the Ising specification with the
additional restriction that there must be a majority of ``$+$'' spins
within each block in $V$.  In an analogous way we define,
 the finite-volume measures
$\pipm_V(\,\cdot\,|\sigma)$,
with the blocks inside $\Lambda$
having a majority of ``$-$''
spins, and those outside a majority of ``$+$''.

We decompose
now  the argument yielding the proof of Claim 3.3
into a sequence
of rather natural observations:

\begin{observation}\label{observation2.1}
There is a unique measure $\mup$ consistent with the
specification $\{\pip_V\}$.  Likewise,
for a fixed finite union of blocks $\Lambda$,  there is a
unique measure $\mupm$ consistent with the
specification $\{\pipm_V\}$.
\end{observation}

Indeed, the uniqueness of $\mup$
(at all temperatures) follows from
ferromagnetic nature of the model and the uniqueness of the
ground state: The latter implies, via Griffiths II
inequality
\cite{gri72}, that for each temperature the expectations
with ``$+$'' boundary conditions are equal to those with
``$-$'' boundary conditions.  This implies uniqueness by
FKG-type arguments \cite{forginkas71}. The uniqueness of
$\mup$ implies that of $\mupm$ because the distributions
$\{\pipm_V\}$ are only a finite-volume modification of the
kernels $\{\pip_V\}$ \cite[Section 7.4]{geo88}.

\begin{observation}\label{observation2.2}
There exists a constant $c$ such that, for $h>J/2$,
\begin{equation}
\pipm_B\Bigl( N_+(B) = 25 \Bigm| - \Bigr)
\;\ge\; 1 - c\, e^{-\beta h}
\label{2.1}
\end{equation}
for any block $B$ outside $\Lambda$.
\end{observation}

This is just the fact that, for $h>J/2$, a block with less than 25
spins ``$+$''
(but with at least 13 pluses)
has,
under minus boundary conditions,  an energy
cost of at least
$\beta h$.  The constant
$c$ is just the number of configurations of such a block,
$c=2^{24}-1$.

\begin{observation}\label{observation2.3}
For each $\delta>0$ there exists a
constant $\widetilde\beta$ such that for
$\beta>\widetilde\beta$ and $h>J/2$
\begin{equation}
\mupm\Bigl( |\scrb(\gamma)|>\delta |\gamma|\Bigr)
\;\le\; \epsilon^{|\gamma|}.
\label{6}
\end{equation}
with $\epsilon<1$, for all families $\gamma$ of
$5\times5$-blocks located {\em outside} $\Lambda$.
\end{observation}

This is proven via the well-known technique of Bernstein's, or
``exponential Chebyshev'', inequality \cite{ber46,ibrlin65}.  To
simplify the notation, let us define a block-random variable
\begin{equation}
X_B \;=\; \left\{\begin{array}{ll}
1 & \hbox{if } N_+(B) < 25\\
0 & \hbox{otherwise.}
\end{array}\right.
\label{2.3}
\end{equation}
We then have
\begin{eqnarray}
\mupm\bigl( |\scrb(\gamma)|>\delta |\gamma|\bigr) &\le&
\mupm\biggl( \ind\Bigl[{\textstyle \sum_{B\in \gamma}} X_B > \delta
|\gamma|\Bigr] \,\exp\Bigl[{\textstyle \sum_{B\in \gamma}} X_B -
\delta |\gamma|\Bigr]\biggr)\nonumber\\
&\le& \mupm\biggl( \exp\Bigl[{\textstyle \sum_{B\in \gamma}} X_B -
\delta |\gamma|\Bigr]\biggr)\;.
\label{2.4}
\end{eqnarray}
(In the first inequality, $\ind[A]$ is the indicator function of the
event $A$.)
By FKG inequalities and Observation \ref{observation2.2},
\begin{eqnarray}
\mupm\biggl( \exp\Bigl[{\textstyle \sum_{B\in \gamma}}
X_B-\delta|\gamma|\Bigr]\biggr) &\le&
\prod_{B\in\gamma} \Bigl[e^{-\delta}\,\, \pipm_B\bigl(
e^{X_B}\bigm|-\bigr) \Bigr]\nonumber\\
&\le& \Bigl[e^{-\delta} \,\bigl( 1 + c\,e^{-\beta h}e \bigr)
\Bigr]^{|\gamma|} \nonumber\\
&\equiv& \epsilon^{|\gamma|}\;.
\label{2.5}
\end{eqnarray}

\begin{observation}\label{observation2.4}
There exists a
constant
 $\beta_2$ such that for $\beta>\beta_2$ and
$h>J/2$ the blocks close to the origin have
$\mupm$-probability larger than 1/2 to be in the
configuration of the ground states of class I (Figure
\ref{f.2}).
\end{observation}

This follows from the preceding observation by a probabilistic Peierls
argument.  Take $\gamma=\partial\Lambda$, that is equal to the blocks
immediately outside $\Lambda$, and $\delta=1/18$.  Then by Observation
\ref{observation2.3} there is a very large probability that the
configuration on
$\partial\Lambda$ look like in Figure \ref{f.3} (a), except for a
small fraction of ``bad'' blocks that does not exceed 1/3rd of the
blocks in the smallest side of $\Lambda$ (because we chose
$\delta=1/18$, see dimensions in Figure \ref{f.3}).  In this
situation, a  standard Peierls argument, as sketched at the
beginning of the proof of the claim, yields the above
observation.  The contribution due to configurations of
$\partial \Lambda$ with a larger fraction of ``bad'' blocks
is bounded by $\epsilon^{|\partial\Lambda|}$ which tends to
zero as $\Lambda$ grows.

\begin{observation}\label{observation2.5}
For any configuration $\sigma$
one has
\begin{equation}
\lim_{V\nearrow\zed^2} \pipm_V(\,\cdot\,|\sigma) \;=\;
\mupm(\,\cdot\,)
\label{2.6}
\end{equation}
(in the weak sense).
\end{observation}

Indeed, every accumulation point of sequences (nets)
$\pipm_{V_n}(\,\cdot\,|\sigma^{(n)})$ is a Gibbs state of the
specification $\{\pipm_V\}$ (it is easy to see that such accumulation
points must satisfy the corresponding DLR equations), but by
Observation \ref{observation2.1} there is only one such a Gibbs state,
namely $\mupm$.
\bigskip

The last observation implies that we can replace $\mupm$ by
$\pipm_{V}(\,\cdot\,|\sigma)$ in Observation \ref{observation2.4}.
This proves Claim \ref{claim2}.
\bigskip

\bigskip

The proof of Theorem \ref{tmaj} can now be completed almost
identically to the proof for block-average transformations in
\cite{vEFS_JSP}:  Claims \ref{claim1} and \ref{claim2} constitute Step
1 and Step 2 respectively, and one can then
proceed to the
 Step 3 (``unfixing'' of
the block spins close to the origin) as in pp.\ 1008-1009 of
\cite{vEFS_JSP}.  The conclusion is that there exists a
sequence of (van Hove) volumes $\Lambda\nearrow\zed^d$ (those shown in
Figure \ref{f.3}) and open sets of (block-spin) configurations ${\cal
N}'_+$ [``$+$'' on an annulus surrounding $\Lambda$ and arbitrary otherwise],
and ${\cal N}'_-$ [``thickened version of those of Figure \ref{f.4}
(b): ``$-$'' immediately above and below $\Lambda$, then
an annulus of ``$+$'' and arbitrary farther out], such that there
exists a constant $c>0$,
{\em independent of}\/ $\Lambda$,
with
\begin{eqnarray}
\lefteqn{\Bigl|
E_{T_L\mu_{\beta, h}}(\sigma'_0+\sigma'_1|\{\sigma'_x\}_{x\neq 0, e_1})
\bigl(-'_\Lambda\eta'\bigr) }\nonumber\\
&& {}-
E_{T_L\mu_{\beta, h}}(\sigma'_0+\sigma'_1|\{\sigma'_x\}_{x\neq 0, e_1})
\bigl(-'_\Lambda\theta'\bigr) \Bigr|
\;>\;c
\label{l5}
\end{eqnarray}
for every $\eta'\in {\cal N}'_+$ and $\theta'\in {\cal N}'_-$.  We have
denoted $e_1=(0,1)$ and $\omega'_\Lambda\eta'$ is the configuration
equal to $\omega'$ inside $\Lambda$ and to $\eta'$ otherwise.  That
is, $T_L\mu_{\beta, h}$ has a conditional probability which is
essentially discontinuous at $\wspec=$``$-$''.  In particular, it can
not be Gibbsian.

\section{Non-Gibbsianness of Decimated Potts Models
\hfill\break
Above the Transition Temperature}

We consider now the $q$-state Potts model in $\zed^d$, which is
defined by spins $\sigma_x\in\{1,\ldots,q\}$ and interaction
\begin{equation}
   \Phi_A(\sigma)   \;=\;
   \cases{
-J(\delta(\sigma_x, \sigma_y)-1)  & if $A = \{x,y\}$
\hbox{ with } x,y \hbox{ nearest neighbors}\cr
            0                          & otherwise $\;$,     \cr
         }
\label{3.1}
\end{equation}
and suppose that
$J>0$.  Here $\delta(\sigma_x,\sigma_y)$
equals 1 if
$\sigma_x=\sigma_y$ and 0 otherwise.
To simplify the notation, we incorporate, in the following,
the coupling $J$ into the inverse temperature $\beta$
(i.e., we put $J=1$ in \reff{3.1}).
 Below we shall also refer
to the corresponding model with a field in the
1-direction.  By that we mean the addition of interaction
terms $h_x
\delta(\sigma_x,1)$ at each
$x\in\zed^d$.

For $q=2$ the Potts model becomes (equivalent to) the Ising model.
On the other hand for large $q$ very different properties emerge, in
particular it is known that for $q$ sufficiently high
the Potts model exhibits a
first-order phase transition \cite{kotshl82,brikurleb85}
with critical inverse temperature
\begin{equation}
\beta_c \;=\; \frac{1}{d} \ln q + O(1/q)\;.
\label{3.0}
\end{equation}
Our results apply to models with $q$ sufficiently high, and
we find it useful to present them in three steps of
increasing technical complication.

\subsection{Lack of Complete Analyticity Above $T_c$}

As a warm-up step we shall show the following:

\begin{theorem}\label{tpotts1}
If $q$ is sufficiently high and the spins
of the sublattice $(N\zed)^d$ are fixed to
be equal to  1,
the resulting system on the rest of the lattice has a first-order
phase transition at a temperature $T_c^{(N)}$ which is strictly larger
than the Potts critical temperature $T_c$.
\end{theorem}

This theorem can be interpreted as showing that at $T_c^{(N)}$ one can
find sequences of volumes (those with ``holes'' at the sites in
$(N\zed)^d$ and boundary conditions
(equal to 1 at the holes and 1 or disordered at
other boundaries)
yielding, in the limit, different one-side derivatives of
the free energy density.
In particular, this means that the
analyticity of the (finite-volume) free energies
cannot be
uniform in the volume and the boundary conditions; that is,
there is no complete analyticity.
\bigskip


We will prove Theorem
\ref{tpotts1}  by transcribing the proof by
Bricmont, Kuroda and Lebowitz
\cite[Theorem 5]{brikurleb85} of the existence of a first-order phase
transition for the regular Potts model.
Before doing so, however, let us briefly show the main
ideas of an alternative proof based on the use of
chessboard estimates.
To minimize technicalities, we will restrict ourselves here
to the case of $N=2$. The proof is particulary simple if
one uses reflection positivity with respect to
(hyper)planes  passing through the sites of the lattice
(see \cite{chakotshl94} for the details of the use of
this particular version of chessboard estimates to the Potts
model). In accordance with the standard use of the method,
one has to evaluate the ``partition functions''$Z^P(T)$
corresponding to the patterns obtained on a torus $T$ by
disseminating, with the help of reflections,  particular
patterns $P$ on a single elementary  (hyper)cube $C$
containing $2^d$ lattice sites. All then boils down to the
verification  of the bounds claiming that the patterns
stemming from  completely disordered configurations on $C$
as well as from the configuration with all spins fixed to
equal 1, are dominating over all remaining patterns.
Recalling that the spins on the sublattice $(2\zed)^d$ are
fixed to equal 1, the first two patterns  yield the
partition functions
$Z^{\hbox{disorder}}(T)\sim
(q^{\frac{2^d-1}{2^d}}e^{-d\beta J})^{|T|}$  and
$Z^1(T)=1$, respectively. For any other
pattern, one easily finds
\begin{equation}
\frac{Z^P(T)}{\max(Z^{\hbox{disorder}}(T),Z^1(T))}\le
\epsilon
\label{a}
\end{equation}
with sufficiently small $\epsilon $.
Indeed, considering for simplicity the two-dimensional
case, we take, as an example, the pattern stemming from the
situation where the horizontal bond attached to  the chosen
site on $(2\zed)^2\cap C$ is ordered and all remaining
(three) bonds in $C$ are disordered. It  yields the
pattern with every horizontal line through sites in
$(2\zed)^2$ ordered (all sites at any such line are set to
equal 1) and with all remaining bonds disordered. As a
result  we are getting
$Z^{P}(T)\sim
(q^{\frac12}e^{-\frac32\beta J})^{|T|}$ and thus \reff{a}
is satisfied for all $\beta$ once $q$ is large enough.
(Namely, we have here $\epsilon = q^{-\frac1{16}}$.)
To show that the transition temperature is asymptotically
behaving like $\beta_c\sim\frac{2^d-1}{2^d}\frac1{d}\log q$,
one has just to notice that it is exactly this value of
$\beta$ for which $Z^{\hbox{disorder}}(T)=Z^1(T)$.
Hence, for large $q$, slightly below $\beta_c$ the
disordered pattern dominates also the ordered one, while
slightly above
$\beta_c$, it is the ordered pattern that is dominating.

Coming back to the proof using the  BKL theory (reviewed
in Section
\ref{smaj}) , we again use the fact  that Theorem
\ref{tpotts1} refers to a Potts model on $\zed^d\setminus
(N\zed)^d$ with a magnetic field in the 1 direction of
strength $h_x=1$ if
$x$ is adjacent to the sublattice
$(N\zed)^d$ and zero otherwise.  One can then choose the ``restricted
ensembles'' $\Omega^D$ and $\Omega^1$ formed respectively by the
disordered and the ``all-1'' configurations:
\begin{eqnarray}
\Omega^D=\Bigl\{\sigma\,\colon\,
\sigma_x\neq\sigma_y \hbox{ for all } &\hspace{-1em}x,y&
\hspace{-1em}\hbox{ nearest
neighbors in } \zed^d\setminus(N\zed)^d
\nonumber\\
&& \hbox{and } \sigma_x\neq 1\hbox{ for }x\hbox{ adjacent to }
(N\zed)^d \Bigr\}\;,
\label{3.10}
\end{eqnarray}
and
\begin{equation}
\Omega^1 \;=\; \{1\},
\label{3.11}
\end{equation}
where $1_x=1$ for all $x\in\zed^d\setminus(N\zed)^d$.
For each of these ensembles one constructs restricted partition
functions, for instance
for any $\omega\in\Omega^D$,
we take
\begin{eqnarray}
Z_R^D(\Lambda,\omega)
&\equiv&
\sum_{\sigma_\Lambda\,\colon\,
\sigma_\Lambda\omega\in\Omega^D}
\exp[-\beta H_\Lambda(\sigma|\omega)]
\nonumber\\
&=& \Bigl|\bigl\{\sigma_\Lambda\in\Omega_\Lambda\,\colon\,
\sigma_\Lambda\omega\in\Omega^D\bigr\}\Bigr|\nonumber
e^{-\beta H_\Lambda^D}\\
&\equiv&
e^{S_\Lambda(\omega)}
e^{-\beta H_\Lambda^D}\;.
\label{3.20}
\end{eqnarray}
The notation of the last line emphasizes the fact that
the term $H_\Lambda(\sigma|\omega)\equiv H_\Lambda^D$ does
not depend on the configurations $\sigma$ and $\omega$
once $\sigma_\Lambda\omega$ belongs to $\Omega^D$ and as a
result we can separate the entropy term
$S_\Lambda(\omega)$.
Notice also that even though, strictly speaking, the
entropy $S_\Lambda(\omega)$ depends on a particular choice
of $\omega\in\Omega^D$, this dependence is asymptotically
negligible [cf. \reff{3.22} below]. On the other hand,
\begin{equation}
Z_R(\Lambda,1) \;\equiv\;
1\;.
\label{3.21}
\end{equation}

The system with restricted ensembles \reff{3.10} and \reff{3.11}, and
restricted partition functions \reff{3.20} and \reff{3.21} satisfies
the requirements (A1)--(A5) of \cite{brikurleb85} just as the usual
Potts model does (p.\ 522--524 of \cite{brikurleb85}).  In particular,
the Peierls condition holds with
\begin{equation}
e^{-\tau} \;\propto\; \frac{1}{q}
\label{sq}
\end{equation}
and the symmetry-breaking parameter is $\beta-\beta_0$, where
$\beta_0$ is the approximate coexistence temperature obtained via
restricted ensembles.  (Hence, $1/q$ plays here the r\^ole that the
temperature plays in the usual Pirogov-Sinai theory, while the
temperature plays the r\^ole of a field).  By the BKL extension of
Pirogov-Sinai theory, we conclude that
there is a temperature where the disordered and ``all-1'' phases
coexist.  Moreover, by \reff{set.6} and \reff{sq}, we have
that, up to corrections of order $1/q$, the transition
temperature is determined by the equality of the restricted free
energy densities, that is by the relation
\begin{equation}
\lim_{\Lambda\,\nearrow\, \zed^d\setminus(N\zed)^d}\,
{S_\Lambda(\omega)\over|\Lambda|} \,\;=\;
\lim_{\Lambda\,\nearrow \,\zed^d\setminus(N\zed)^d}
{\beta H_\Lambda^D}\;.
\label{3.22}
\end{equation}
The limiting value of the left hand side  in \ref{3.22}
actually does not depend on a partricular choice of
$\omega\in\Omega^D$. To construct a disordered
configuration, the number of choices per site is at least
$q-2d$ (assuming all the neighboring spins have been
chosen), and at most
$q$.  Hence,
\begin{equation}
S_\Lambda(\omega) \;=\; |\Lambda|\, \bigl[\ln q +
O(1/q)\bigr]\;.
\label{3.23}
\end{equation}
On the other hand,
\begin{equation}
H_\Lambda^D \;=\; |\Lambda| \,d
\left(1+\frac{1}{N^d-1}\right) +O(|\partial\Lambda|)\;,
\label{3.24}
\end{equation}
where the term
 $d|\Lambda|/(N^d-1)$ is due to the interaction between
spins in $\Lambda$ and spins on the decimated sublattice
$\zed^d\setminus(N\zed)^d$.  From \reff{3.22}--\reff{3.24} we get
\begin{equation}
\beta_c^{(N)} \;=\; {N^d-1\over N^d} \, {1\over d} \ln q
+O(1/q)\;,
\label{3.25}
\end{equation}
which, for large $q$, is smaller, by a factor $(N^d-1)/N^d$,
than the Potts inverse critical temperature \reff{3.0}.

\subsection{Non-Gibbsianness for a Sequence of Temperatures
\hspace*{\fill}\break
Above $T_c$}

Theorem \ref{tpotts1} amounts to proving what in
\cite{vEFS_JSP}  (see eg.\ p.\ 990) was referred to
as Step 1 of the proof of non-Gibbsianness (more precisely,
non-quasilocality) of the renormalized measure.  Such a version
of Step 1, however, can not be extended to a full proof of
non-Gibbsianness because $\wspec$ is a ``maximal'' block-spin
configuration, and hence there is no way to select the different
(internal-spin) pure phases just via block-spin boundary
configurations (that is, Step 2 fails).  This type of difficulty is
already present in other expected examples of non-Gibbsianness
proposed in the literature (see discussion in pp.\ 1006--1007 of
\cite{vEFS_JSP}).

To circumvent this problem, one must prove the analogue of Theorem
\reff{tpotts1} but
for decimated spins fixed in some non-uniform
configuration.  This is easily accomplished:  take a periodic
configuration in $\zed^d\setminus(N\zed)^d$ with a fraction
$f <1/2$ of spins chosen
to equal  2 and the rest to equal  1.  The
same arguments as in the previous section apply, except that
\reff{3.24} is generalized to
\begin{equation}
H_\Lambda^D \;=\; |\Lambda| \,d \left(1+
\frac{1-2f }{N^d-1}\right)
+O(|\partial\Lambda|)\;,
\label{3.26}
\end{equation}
hence the coexistence between the ``all-1'' and disordered phases
takes place at an inverse temperature
\begin{equation}
\beta_c^{(N,f )} \;=\; {N^d-1\over N^d-2f } \,
{1\over d} \ln q +O(1/q)\;,
\label{3.27}
\end{equation}

As a result, we now have two phases that can be selected via
decimated-spin boundary conditions:  if such spins are chosen to be 1
then the ``all-1'' phase is singled-out; and any choice disfavoring
it, for instance boundary decimated spins 3, selects the disordered
phase (Step 2 of \cite{vEFS_JSP}).  The argument can be completed as
for decimation of Ising spins (Step 3 in \cite{vEFS_JSP}) to prove the
discontinuity of the decimated conditional probabilities at the
inverse temperatures $\beta_c^{(N,f )}<\beta_c$.  We notice that
for fixed $N$ (decimation scheme), these inverse temperatures
range from $\beta_c^{(N)}$ of the previous section (for $f =0$) and
the Potts model $\beta_c$ given in \reff{3.0} (for $f =1/2$).  As
discussed in the previous section, our proof of non-Gibbsianness does
not apply for $f =0$.  It does, however, apply at $f =1/2$ where at
the corresponding critical temperature there are {\em three}\/
coexisting phases:  ``all-1'', ``all-2'' and disordered.

On the other hand, the term ``$O(1/q)$'' in \reff{3.27} is {\em not}
uniform in the period of the decimated configuration chosen.  In fact,
a closer look
at the proof of Bricmont, Kuroda and Lebowitz reveals
that the larger the period, the larger the minimal value of $q$
needed.  Hence, for each fixed $q$ (and $N$), there is only a finite
set of qualifying fractions $f $, that is, the argument yields only
a {\em finite}\/ sequence of critical inverse temperatures.

We summarize the results of this section:
\begin{theorem}\label{tpotts2}
For each dimension $d\ge2$ and each decimation of period $N$ there
exists a $q_0$ such that for each $q>q_0$ there exists a finite
sequence of temperatures $\{T_c^{(N,f(q))}\}$, $f (q)$ taking
finitely many values in $\bbbq\cap(0,1/2]$, larger than
the Potts critical temperature, for which the measure arising by
decimation of the $q$-Potts model is not
consistent with any quasilocal specification,
 in particular, it is not
Gibbsian.
\end{theorem}

\subsection{Non-Gibbsianness for an Interval of Temperatures
\hspace*{\fill}\break
Above $T_c$ ($d\ge3$)}

The limitations of the method of the previous section (finite sequence
of particular temperatures) can be overcome by choosing the
decimated spins in a {\em random}\/ fashion, for instance
2 with probability $f $ and 1 otherwise.  By using a random version
of Pirogov-Sinai due to Zahradn\'{\i}k \cite{zah93} we can then prove
the analogue of Theorem \ref{tpotts2} for a whole interval of
temperatures above $T_c$.  Zahradn\'{\i}k's proof of the
existence of coexisting phases for random systems only applies for
small disorder ($f $ small) and dimensions $d\ge3$.

This part of the argument is technically
complicated, but is essentially identical to the one given in
\cite[pp.\ 1012--1013]{vEFS_JSP} for the Ising model, except that for
Potts models $1/q$ plays the r\^ole of the temperature in
low-temperature
Ising models and the temperature plays the r\^ole of the magnetic
field.  We opt for skipping the details and content ourselves with
stating the conclusions.

\begin{theorem}\label{tpotts3}
For each dimension $d\ge3$, and each decimation period $N$ there
exists a $q_0$ such that for each $q>q_0$ there exists a non-empty
interval of temperatures $\bigl(T_c,T(q)\bigr)$ where the measure arising from
the decimation of the $q$-Potts model is not
consistent with any quasilocal specification, in particular it is not
Gibbsian.  The temperatures $T(q)$ increase with $q$.
\end{theorem}

\section{Conclusions and Final Comments}
\label{sconc}

We have shown examples of renormalization transformations exhibiting
pathologies deep inside the one-phase region and (for the first
time) within the high-temperature phase.  These examples suggest that
the occurrence of this type of pathologies is a rather robust
phenomenon.  It is still not clear, however, what the practical
consequences of these pathologies are.

A natural question is the size of the set of ``pathological''
configurations $\wspec$ at which some finite-volume conditional
probability is non-quasilocal (discontinuous).
In the case of the majority-rule acting on the Ising model in a strong
field, this set of pathological configurations is of measure zero with
respect to the (unique) Ising Gibbs state.  This follows from the
results of \cite{ferpfi94}.  The same is true for the case of block
averaging in a field (analyzed in \cite[p.\ 1014]{vEFS_JSP}).  This
raises the possibility of restoring a weak form of Gibbsianness
defined only almost-surely \cite{benmaroli94,lor94,lorwin92,dobshlp,hay94}.

For the high-temperature pathologies of the decimated Potts models, we
expect them to disappear if the decimation transformation is repeated
sufficiently many times.  Alternatively, for any temperature above
$T_c$ the pathologies should be absent if the decimation is taken with
linear period $N$ large enough.  This expectation is based on similar
results obtained by Martinelli and Olivieri \cite{maroli93} for the
Ising model in nonzero field (which is the analogue of $T>T_c$ for the
Potts-model transition).  On the other hand, for any fixed $N$
our Theorem \ref{tpotts3}
implies that for $q$ large enough every open interval around the
transition temperature $T_c$ includes (a whole subinterval of)
temperatures where the decimation transformation produces
non-Gibbsianness.  This is to be contrasted with some results
\cite{ken92,benmaroli94,vel94} suggesting an opposite conclusion for
neighborhoods of the critical temperature of the Ising model.
Although the arguments presented in these works are not completely
rigorous ---~they are based on numerical studies of a small number of
decimated configurations~---
one may indeed expect differences between
the cases for which
there is a continuous phase transition at $T_c$
(low-$q$ Potts models) and
the cases where the phase transition at $T_c$
is of first order (the high-$q$ Potts models analyzed here).

\section*{Acknowledgments}
A.C.D.v.E.\ and R.F.\ have developed most of their insights into
non-Gibbsianness in collaboration with Alan D.\ Sokal.  Furthermore,
we have had very useful discussions with R.L.\ Dobrushin, J.\
L\"orinczi, E.\ Olivieri, C.-Ed.\ Pfister, S.B.\ Shlosman, M.\ Winnink
and M.~Zahradn{\'\i}k.
R.F.\ thanks the Institute of Theoretical Physics of the R.U.\
Groningen for hospitality while this paper was being written.
The research of the first author (A.C.D.v.E.)\ has been made possible
by a fellowship of the Royal Netherlands Academy of Arts and Sciences (KNAW).
The research of the second
author (R.F.)\ was supported in part by
the Fonds National Suisse
and the third author (R.K.)\ by the grants GA\v CR
202/93/0449 and GAUK 376.

\addcontentsline{toc}{section}{\bf References}

\end{document}